\documentstyle[12pt,pre,aps,floats,epsf,amssymb]{revtex}
\draft

\title{Cross-linking of micelles by gemini surfactants}
\author{Prabal K. Maiti$^{1,*}$ and  Kurt Kremer}
\address{Max Planck Institute for Polymer Research,\\
Postfach 3148, 55021 Mainz, Germany}
\author{Oliver Flimm, Debashish Chowdhury$^{\dagger}$ and Dietrich Stauffer}
\address{Institute for Theoretical Physics,\\
University of Cologne, 50923 K\"oln, Germany}
\begin{document}
\maketitle

\begin{abstract}
We investigate the effects of gemini surfactants, telechelic chain
and lipids on the nature of micelles formed by conventional 
single-tail surfactants in water by carrying out Monte Carlo 
simulations. In a mixture of gemini and single-tail surfactants in 
water we find direct evidence of micelles of predominantly 
single-tail surfactants some of which are dynamically cross-linked 
by gemini surfactants when the concentrations of the geminis is 
only a few mole percent and their spacers are {\it hydrophilic}. 
In contrast, mixtures of lipids and single-tail surfactants in water 
form only isolated micelles, each consisting of a mixture of both 
species, without cross-links. 
\end{abstract}
{\bf Running title:} Mixed and cross-linked micelles

\newpage

\section{Introduction:}

Amphiphilic molecules are not only important constituents of 
bio-materials, but also find wide-ranging industrial applications 
~\cite{Evans}.  These surface-active agents are also referred to 
as surfactants.  Many surfactants consist of a single "hydrophilic 
head" connected to a single "hydrophobic tail" whereas some other 
surfactants, e.g., phospholipids, are made of two hydrophobic tails 
both of which are connected to the same hydrophilic head (fig.
\ref{molecule}). In contrast, {\it gemini} surfactants 
~\cite{Deinega,Menger1,Menger2,Rosen1,Zana1,Zana2,Zanarev} 
consist of two single-tail  surfactants whose heads are connected 
by a "spacer" chain (fig.\ref{molecule}). The spacer in gemini 
surfactants may be hydrophilic or hydrophobic \cite{Rosen2}. 
Because of their "water-loving" heads and "water-hating" tails, 
surfactant molecules form "{\it self-assemblies}"  (i.e., 
supra-molecular aggregates), such as micelles, vesicles, etc., 
\cite{Gelbert} in aqueous media. For example, micelles are formed 
when the concentration of the surfactants in water exceeds what 
is known as the "critical" micellar concentration (CMC). Some of 
the unusual properties of surfactants, which are exploited in their 
wide-ranging industrial applications, are crucially influenced by 
the nature of their aggregation in aqueous environments.

In this paper we address the following question: what are the
effects of a few mole percent of gemini surfactants, mixed with
single-tail surfactants, on the nature of micellar aggregates
in water? By carrying out Monte Carlo (MC) simulations of a 
mixture of gemini and single-tail surfactants in water we show 
that in such a system, in equilibrium, micelles of predominantly 
single-tail surfactants are physically (i.e., non-covalently) 
"cross-linked" by geminis when the spacers are hydrophilic and 
the relative concentration of the geminis is only a few mole 
percent. The direct evidence emerging from the instantaneous 
snapshots of the system in our simulations may settle a controversy 
as to the interpretation of indirect experimental indications
\cite{meli,zanal1,zanal2}. Moreover, we find lipids (i.e., 
double-tail surfactants) fail to crosslink micelles of single-tail 
surfactants; instead, isolated micelles, each consisting of a 
mixture of lipids and single-tail surfactants, form spontaneously 
in water. Furthermore, we also show that "telechelic chain" 
\cite{Khalatur,Semenov}, another class of amphiphilic molecules, 
succeeds less often than the geminis in cross-linking micelles. 

\section{The model} 

In the spirit of lattice gas models, we model the fluid under 
investigation as a simple cubic lattice of size 
$L_x \times L_y \times L_z$. A water molecule occupies a single 
lattice site. The surfactants are modelled as short chains on the 
lattice. The "primary structure" of a single-chain surfactant is 
defined by the symbol\cite{Stauffer,stau1,Livrev} 
${\cal T}_m{\cal N}_p{\cal H}_q$ 
where ${\cal T}$ denotes {\it tail}, ${\cal H}$ denotes {\it head} 
and ${\cal N}$ denotes the 'liaison' or neutral part of the 
surfactants. The integers $m$, $p$ and $q$ denote the lengths of 
the corresponding pieces, giving a chain of length $\ell = (m+p+q)$.
Bernardes \cite{bernardes} extended this to a microscopic lattice 
model of lipids in water. The "primary structure" of lattice model 
of lipids can be described by the symbol 
${\cal T}_m{\cal N}_p{\cal H}_q{\cal N}_p{\cal T}_m$. In terms of 
the same symbols, the gemini surfactant can be represented \cite{maiti}
by the symbol 
${\cal T}_m{\cal N}_p{\cal H}_q{\cal S}_n{\cal H}_q{\cal N}_p{\cal T}_m$
where $n$ is the number of lattice sites constituting the spacer 
represented by the symbol ${\cal S}$. Finally, a microscopic 
lattice model of "telechelic" chains can be represented by the 
symbol ${\cal T}_1{\cal N}_p{\cal H}_n{\cal N}_p{\cal T}_1$. 
In other words, a telechelic chain can be viewed as a gemini 
surfactant with hydrophilic spacer and the shortest possible 
hydrophobic tails.

Using these lattice models of surfactants in water, in this 
paper we investigate the effects of small mole fractions of 
geminis (or lipids or telechelic chains) on the micellar aggregates 
formed by the single-tail surfactants. We carry out Monte Carlo 
(MC) simulations of a mixture of single-tail and gemini surfactants 
in water for $p = q =1$ and for tail length $m=4$ to $m=12$ and 
spacer length $n = 0$ to $n=14$.  We shall refer to each site on 
the surfactants as a {\it monomer}. Note that in this type of 
models \cite{smit} the "water-loving" head group is assumed to 
be "water-like". 

The above mentioned model of single-chain surfactants in aqueous 
media \cite{Lar1} has been reformulated \cite{Stauffer,stau1} 
in terms of Ising-like variables, in the same spirit in which a 
large number of simpler lattice models had been formulated earlier 
\cite{Gomsch} for the convenience of calculations. In this 
formulation, a classical Ising spin-like variable $S$ is assigned 
to each lattice site; $S_i = 1$ ($-1$) if the $i$-th lattice site 
is occupied by a water (oil) molecule. If the $j$-th site is 
occupied by a monomer belonging to a surfactant then 
$S_j = 1, -1, 0$ depending on whether the monomer at the $j$th 
site belongs to head, tail or neutral part (see fig.\ref{model}). 
The monomer-monomer interactions are taken into account in analogy 
with the ferromagnetic nearest-neighbour interaction in Ising magnets. 
Thus, 
the thermal Boltzmann probability for 
a configuration is proportional to $\exp(-H/k_BT)$ and 
the Hamiltonian for the system is given by the
standard form
\begin{equation}
H = - J \sum_{<ij>} S_i S_j.
\end{equation}
The temperature $T$ of the system is measured in the units of $J/k_B$ 
where $J$ denotes the strength of the interactions between the 
spin-like variables $S$ on nearest-neighbour sites and $k_B$ is the 
Boltzmann constant (while varying $T$ we assume $J$ to be constant).  
Moreover, in order to investigate the role of 
the chain stiffness we have used a chain bending energy~\cite{Chow2}; 
every bend of a spacer, by a right angle at a lattice site, is assumed 
to cost an extra amount of energy $K (>0)$.

This type of microscopic lattice models is very useful for 
investigating the generic qualitative features of a class of 
surfactants rather than for a quantitative prediction of any 
dynamic property of any specific material; for the latter 
purpose molecular dynamics simulations of molecular model of 
the specific surfactant is more appropriate, although much 
more computer time consuming.

In order to investigate the spontaneous formation and morphology 
of supra-molecular aggregates, we initially disperse the model 
gemini surfactants and single tail surfactants randomly in a 
$L_x \times L_y \times L_z$ system which contains only water other 
than the surfactants. Since we are interested in the effects of 
gemini surfactants on micellar aggregates formed by single-chain 
surfactants the concentration of the single tail surfactants in 
all our simulations is well above its characteristic micellar 
concentration (CMC) while the concentration of the gemini surfactants 
is much lower. The total volume fraction $\phi$ is defined by
$\phi = (N_{st} \ell + N_{gem}[2\ell+n])/(L_x L_y L_z)$
where $N_{st}$ and $N_{gem}$ are the total numbers of single-tail
and gemini surfactants, respectively, in the system. Moreover,
the relative volume fractions of the geminis and the single-tail
surfactants are defined by
$c_{gem} = (N_{gem} [2\ell+n])/\biggl(N_{st} \ell + N_{gem} [2
\ell+n]\biggr)$

We allow the same moves of the surfactants and water molecules 
as those allowed in ref.\cite{Livrev,bernardes,maiti}. 
The moves allowed for the surfactants in our model are as
follows:\\
(i) {\it reptation:} this is identical to the reptation move for single-chain
surfactants 
(also called "slithering snake" move) \cite{Livrev,bernardes,maiti}; 
(ii) {\it spontaneous chain buckling:} a portion
in the middle of one of the two tails or the spacer is randomly picked up
and allowed to buckle with the 
Boltzmann probability mentioned above; 
(iii) {\it kink
movement:} a kink formed by the buckling or reptation is allowed to move
to a new position with the appropriate Boltzmann probability; 
(iv) {\it pull move:} this is the
reverse of spontaneous chain buckling; a buckled part of one of the two
tails or the spacer is pulled so as to make it more extended. 
Each of these moves is possible only
if the new positions of all the monomers are not occupied simultaneously
by monomers belonging to other surfactants. Each surfactant is allowed
to try each of the above mentioned moves once during each MC step.
The system evolves following the standard Metropolis algorithm 
\cite{binder} while we monitor the mean-size of the clusters of 
the surfactants. After sufficiently long time, if the mean size of 
the clusters of the surfactants attains a time-independent value 
it indicates that the system has been equilibrated. 
All the results (except the snapshots) reported in this paper have 
been generated for system sizes
$L_x = L_y = L_z = L = 100$ by averaging over sufficiently large (10-25)
number of runs.

We have defined CMC as the amphiphile concentration where half of 
the surfactants are in the form of isolated chains and the other 
half in the form of clusters consisting of more than one neighbouring 
amphiphile \cite{Stauffer,maiti}.

To quantify the process of cross-linking we have calculated the number 
of cross-links in the system as follows: 
a) First we make a Hoshen-Kopelman~\cite{hoshen} cluster count without 
the gemini surfactants i.e we identify and mark the cluster formed by 
the single tail surfactants.
b) Next we do a second Hoshen-Kopelman cluster count with both gemini 
and single tail surfactants.
c) When a cluster in (b) dissolves in two or more clusters in (a) we 
have found a cross-linked micelle.
The number of surfactant monomers per micelle is also counted by a 
Hoshen-Kopelman algorithm \cite{hoshen,flimm}.

\section{Aggregation Numbers}

We have computed the aggregation numbers of micelles present in the 
gemini/single tail surfactants at different temperature and for 
several concentrations. Figure \ref{gagg} and \ref{sagg} show the 
variations of the aggregation number of the pure non-ionic gemini
and single tail surfactants for different concentrations and 
temperatures. The aggregation number of the pure gemini micelles
increases with concentration and temperature as for conventional 
non-ionic surfactants. For the single
tail surfactants (${\cal T}_{10}{\cal N}_1{\cal H}_1$ )
the mean aggregation number has the same behaviour
as that of gemini surfactants. At a given temperature the aggregation
number is higher for higher concentration of surfactants. This is in 
qualitative agreement with the previous experimental results 
~\cite{zanaold}. Figure ~\ref{mixagg} shows the variation of $N_A$, 
the total number of surfactant monomers (gemini plus single tail) 
per micelle in the mixed system, at a fixed total surfactant 
concentration, as a function of the percentage of the gemini surfactant 
present in the mixture. It is evident form figure ~\ref{mixagg} that 
the addition of the gemini surfactants swells the micelle size.

\section{Cross-Linked Micelles}

Addition of the gemini surfactants have a profound effect on the 
mechanism of micellization and overall micellar morphology.
A few of the different types of supra-molecular aggregates that 
are expected to be formed by a mixture of single-tail surfactants 
and geminis in water are shown schematically in fig.\ref{schematic}. 
It is possible that the gemini bind in an intramicellar fashion 
shown in mixed micelle (figure 6(a)) and the spacer reside at the 
micelle surface. The micelles may also be cross-linked (figure 6(b)). 
Another possibility is that the gemini reside in the middle of the 
micellar aggregates. 

An instantaneous snapshot of two spherical micelles cross-linked 
by a gemini surfactant is shown in fig.\ref{dumbbell}; the two 
micelles consist, almost exclusively, of single-tail surfactants 
while the cross-linking is done by a gemini surfactant. However, an 
instantaneous snapshot of a more exotic supra-molecular aggregate 
is shown in fig.\ref{necklace}; a "necklace"-like aggregate (with 
branch) is formed by the cross-linking of a sequence of micelles 
by gemini surfactants. These snapshots not only provide the most 
direct evidence in favour of cross-linking of micelles but also 
establish the formation of "aggregates of micellar aggregates" by 
the mixture of single-tail surfactants and geminis in water.  

In our simulations of a mixture of "telechelic chain" and single-tail 
surfactants also, we get cross-linked micelles. However, the number 
of cross-links is less than that in the case when gemini surfactant 
is present. When a "telechelic chain" cross-links, the only gain in 
energy comes from the fact that their heads find themselves close 
to other heads of single-chain surfactants on the surface of the 
micelles. But, they do not have any tails to hide and therefore, not 
as much urge to cross-link as the gemeinis. 

The cross-linking of the micelles is dynamic, rather than permanent, 
in the sense that such links can break while new links can form 
elsewhere in the system with time.  To our knowledge this is the 
first direct evidence of the formation of cross-linked micelles. 
From our extensive simulations we find that they occur, for both 
flexible as well as rigid hydrophilic spacer, when the spacer is  
sufficiently long. When the spacer is hydrophilic there is the 
following scenario (figure \ref{schematic}):
\begin{itemize}
\item If $d_{mic} \gg $ typical
extension of the spacer, case 6(a) is preferred. In that case we get
swelling of micelles. 
\item If $d_{mic} \leq $ typical extension of the spacer, both cases 
6(a) and 6(b) preferred.
\item If $r_{mic} \ll $ typical extension of the spacer then 
cross-linking is preferred. For long spacer the correction in the 
entropic term is logarithmic, the cross-linking is weakly preferred.
\end{itemize}

Figure ~\ref{cross} shows the mean number of cross-linked micelles 
as a function of the fraction of the gemini surfactants present in the
mixture. At high gemini concentration geminis form their own micelles
rather than cross-linking. By forming their own micelles geminis gain 
entropically. That is why the number of cross-linked micelles decreases
when there is large percentage of geminis present in the system.

We have also computed the CMC for the mixtures of geminis and single 
tail surfactants. In Figure ~\ref{cmc} we have plotted the CMC of the 
mixture against the fraction of gemini surfactants. The plot shows 
the presence of a minimum which indicates synergism in micelle 
formation in the mixture. Synergism in mixed micelle formation has 
been previously reported ~\cite{Rgao,Rzhu} for mixture of anionic 
gemini surfactants and various conventional single tail non-ionic 
surfactants.

We have also carried out simulations when the spacer is rigid.
One of the very surprising observations is that when we have
rigid spacer, we get crosslinking both for hydrophilic as well 
as hydrophobic spacers. However for hydrophobic spacers we 
observed fewer cross-linked micelles. Figure \ref{bendcross} shows 
a snapshot of the cross-linked micelle when the rigid spacer is 
hydrophobic. This is in qualitative agreement with the result of 
Menger et. al. ~\cite{meli} where they postulated the occurence 
of micelle cross-linking in a mixture of CTAB and gemini surfactant 
with rigid spacer. 

We have also carried out simulations when the length of the 
spacer $n=0$; this corresponds to model lipids rather than 
geminis (see fig. 2). In this case, however, we do not find 
any cross-linked micelles. Instead, we observed {\it mixed 
micelles} which have been shown schematically in fig.\ref{schematic}.
Hydrophilic spacers gain energy by remaining surrounded by water.
That is why in the snapshots of micellar aggregates we see that
a larger number of monomers belonging to the spacers are in contact
with water. We believe that the overall loss of entropy caused by
the formation of "cross-linked" micelles cannot be compensated by
the gain of energy if the hydrophilic spacers are not long enough.
This explains why mixtures of single-tail surfactants and lipids
were never seen to form cross-linked micelles in our simulations.

We have also studied quantitatively the growth of the micelles with 
time starting from initial states where surfactants are dispersed 
randomly in water. Initially, we observe an increase of 
the mean-cluster size with the passage of time (see fig.\ref{growth}); 
this indicates the spontaneous formation and subsequent growth of the 
micellar aggregates. Interestingly, the mean-cluster size appears 
not only to follow a power-law growth but also displays three 
different regimes of growth, each characterized by a distinct 
growth exponent (fig.{\ref{growth}) for extremely long times. 
For higher temperatures the growth stopped after some time.

The formation and growth of micelles from a random initial configuration 
take place during the first regime ($0 - 14$ million MCS). In the 
second regime ($14 -42$ million MCS) the micelles start coalescing 
thereby forming an intermediate structure and this process is very fast. 
During the next regime ($42 - 200 $ million MCS) this intermediate 
structure spontaneously forms vesicles which get cross-linked with 
either vesicles or micelles. Thus, from this unusually long runs we 
find a transition from micelles to vesicles. A snapshot showing the
vesicles cross-linked with micelles is shown in figure \ref{vestrans}.
Figure \ref{vesslice} shows a cross-section of the vesicle.
Our simulations might provide some insight into the kinetics
and possible existence as well as structure of intermediates during the
micelle-to-vesicles transition. Although the exponents are 
close to those associated with Ostwald ripening in binary 
alloys and spinodal decomposition in binary fluids \cite{gunton}, 
we leave it for future publications to explore this non-equilibrium 
growth process.

\section{Discussion}

Here we try to make some connection between the available experimental 
parameters and our simulation parameters. For the snapshot shown in 
figure ~\ref{dumbbell} the number of gemini surfactants is $32$
and the number of single tail surfactant is $1200$. Since the total 
lattice site is $10^6$, the mole fraction 
is $3.2 \times 10^{-5}$ for geminis and $1.2\times 10^{-3}$ for
single-tail surfactants.
These mole fractions are in the same range of experimental mole fractions 
reported by Menger et. al ~\cite{meli} where they postulated the
possibility of micelle cross-linking. In their experiments the mole fraction
of the gemini was in the range $1.9\times 10^{-5}$ to 
$5.6\times 10^{-4}$ and that for CTAB
(single tail surfactants used in the experiment) was in the range
$5.0\times 10^{-3}$. Moreover, CMC of CTAB is $8\times 10^{-4}$.
The CMC of the single tail surfactants ${\cal T}_4{\cal N}_1{\cal H}_1$ 
used in our simulation
is in the range $10^{-3}$ to $10^{-4}$ ~\cite{stau1} which is 
comparable to the CMC of CTAB.
These comparisons suggest that our choice of parameters is reasonable and 
our simulation results can be compared with the corresponding experimental 
results. 

The present study gives direct evidence of the formation of cross-linked 
micelles in a mixture of single tail and gemini surfactants. When 
the spacer of the gemini is hydrophilic and sufficiently long cross-linking 
occurs. However when the spacer is hydrophobic, cross-linking occurs only
when the spacer is completely rigid. 
Our simulation also elucidate the mechanism  of mixed micellization.
We have also calculated 
the aggregation numbers of the mixed micelles as a function of 
temperature as well as concentration. The variation of the CMC of the 
mixture shown synergism in micelle formation. From our longest simulation
run we observe a micelle-vesicle transition. This might give us 
some understanding of the growth process and intermediate structure
during micelle-vesicle transition. Our preliminary results in a mixture
of telechelic chain and single tail surfactants also gives rise 
to cross-linking of micelles. Finally our simulations show how the 
the micellar size and its morphology can be controlled by adding 
dimeric surfactants or telechelic chains.

{\bf Acknowledgements:} We thank R. Zana for useful comments and 
B. D\"unweg for helpfull discussions. PKM and DC acknowledge financial 
supports from the Alexander von Humboldt foundation and SFB $341$ 
K\"oln-Aachen-J\"ulich, respectively.

\newpage
%
\begin{figure}[h]
\epsfxsize=\columnwidth\epsfbox{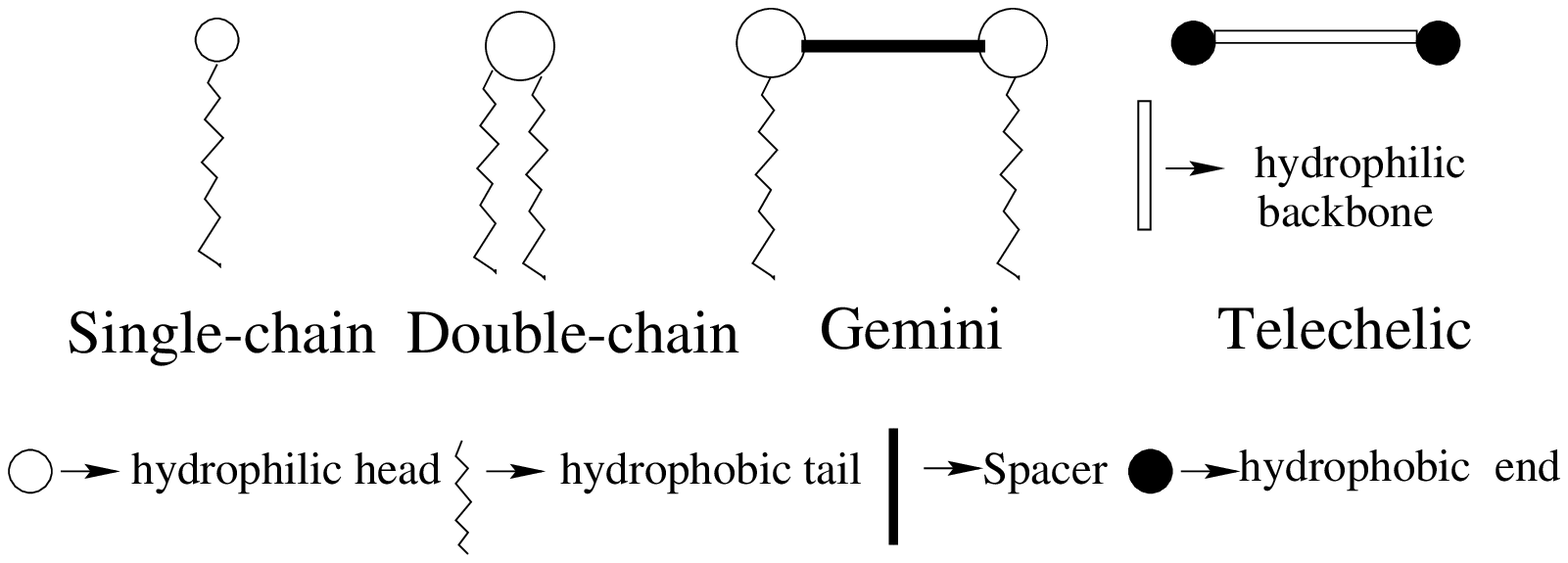}
\caption{\protect{ A schematic representation of different types of
amphiphiles. }}
\label{molecule}
\end{figure}

\begin{figure}
\epsfxsize=\columnwidth\epsfbox{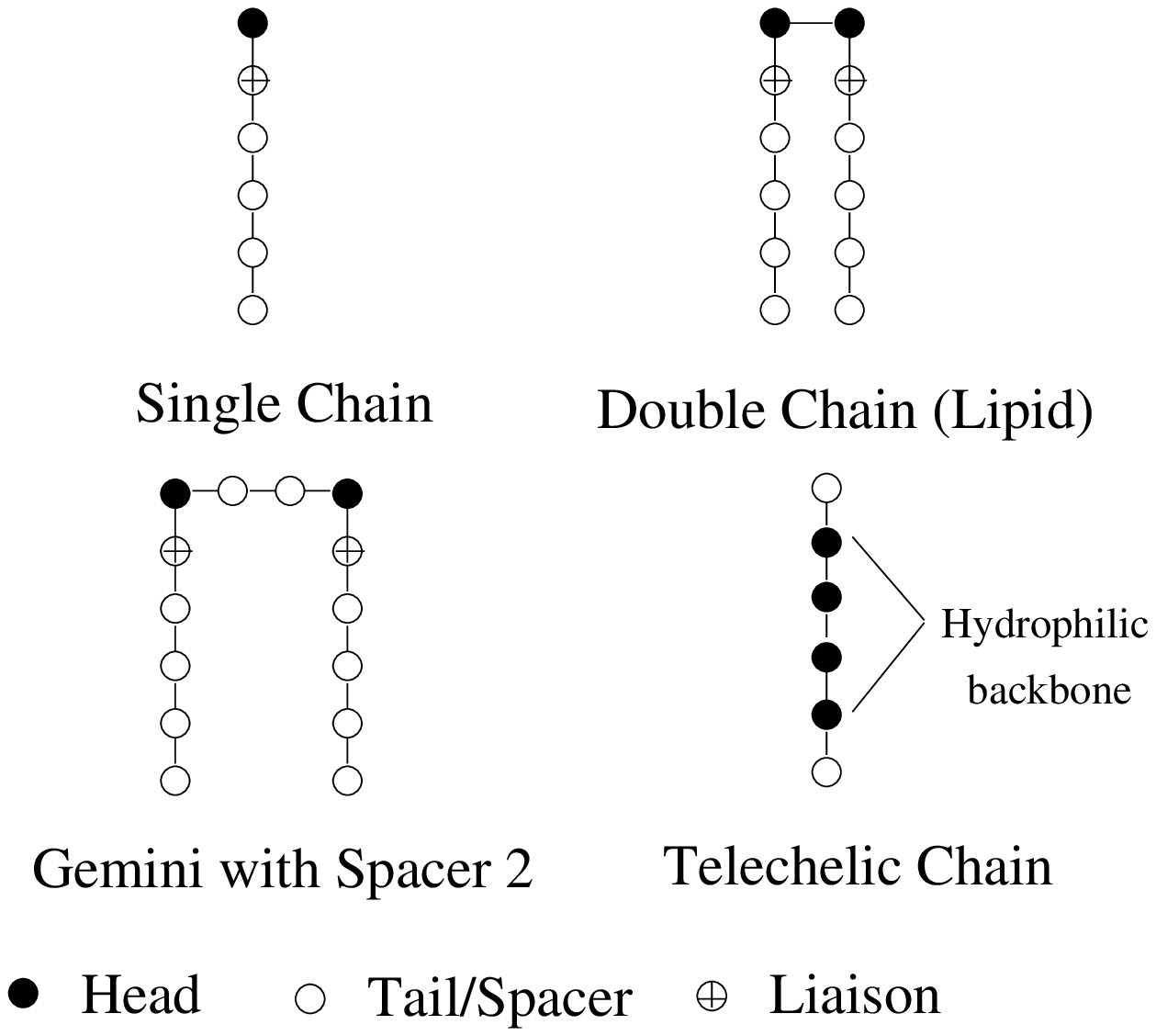}
\caption{\sf Models of single-chain, double-chain, gemini surfactants and
telechelic chain.}
\label{model}
\end{figure}

\begin{figure}
\epsfxsize=\columnwidth\epsfbox{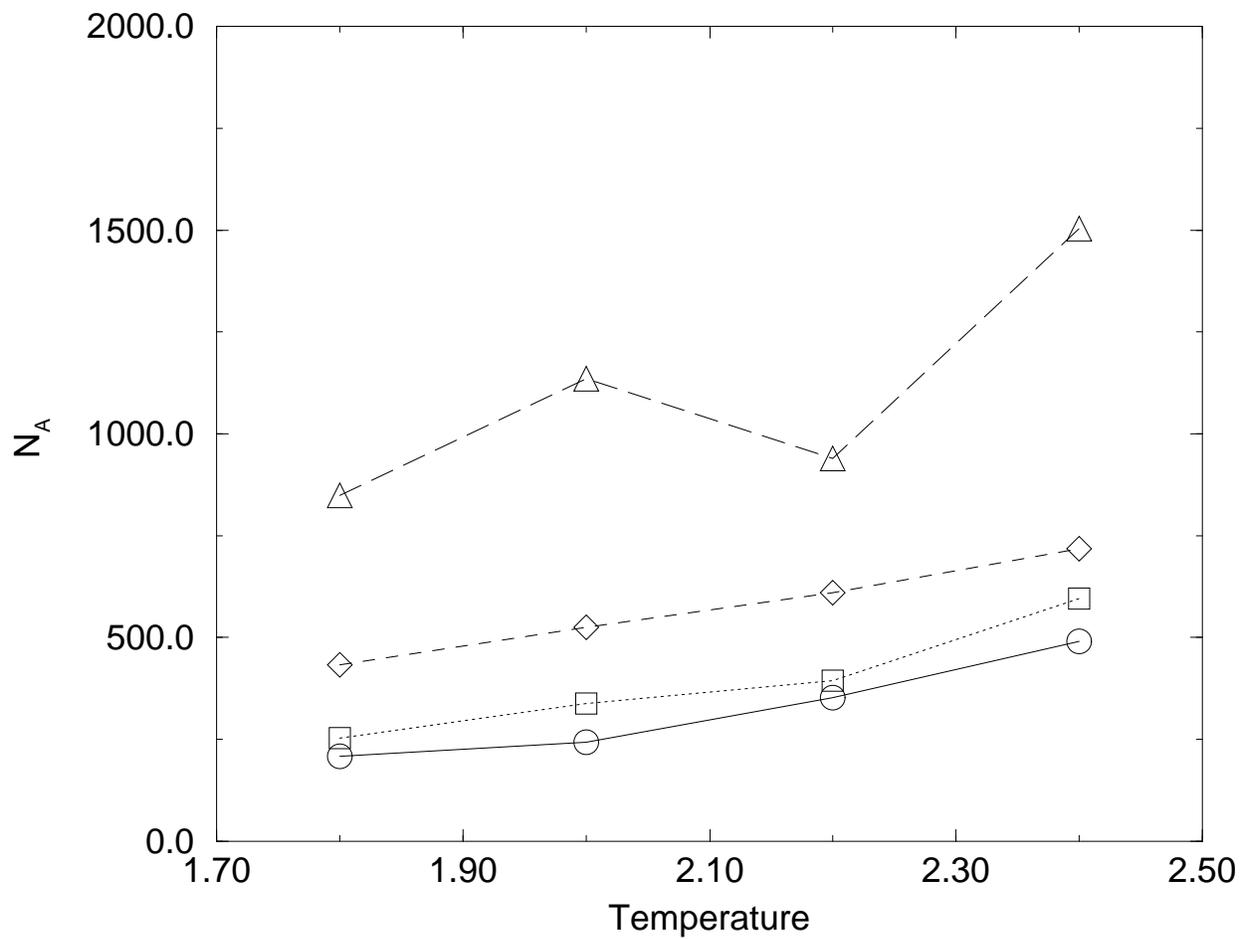}
\caption{\sf Variation of micelle aggregation number $N_A$ with temperature
formed by gemini surfactants in water for gemini concentration
$c_{gem} = 0.005 (\circ)$; $0.01 (\square)$; $0.02 (\diamond)$; and $0.03
(\triangle)$.}
\label{gagg}
\end{figure}

\begin{figure}
\epsfxsize=\columnwidth\epsfbox{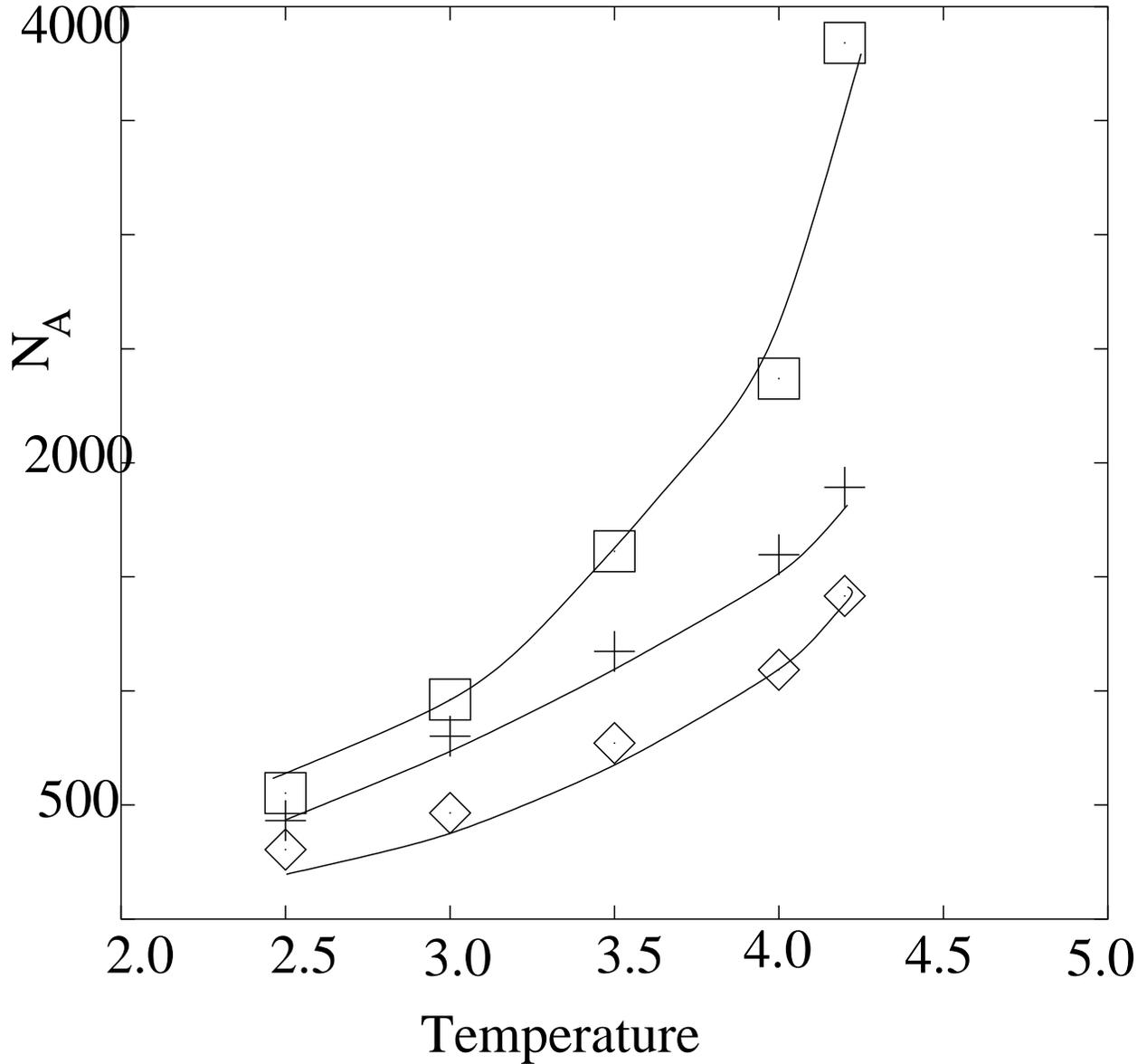}
\caption{\sf Variation of micelle aggregation number $N_A$ with temperature
formed by single-tail surfactants in water for surfactant concentration
$c_{st} = 0.005 (\diamond)$; $0.013 (+)$; and $0.022 (\square)$. The continuous
lines are merely guide to the eye.}
\label{sagg}
\end{figure}

\begin{figure}
\epsfxsize=\columnwidth\epsfbox{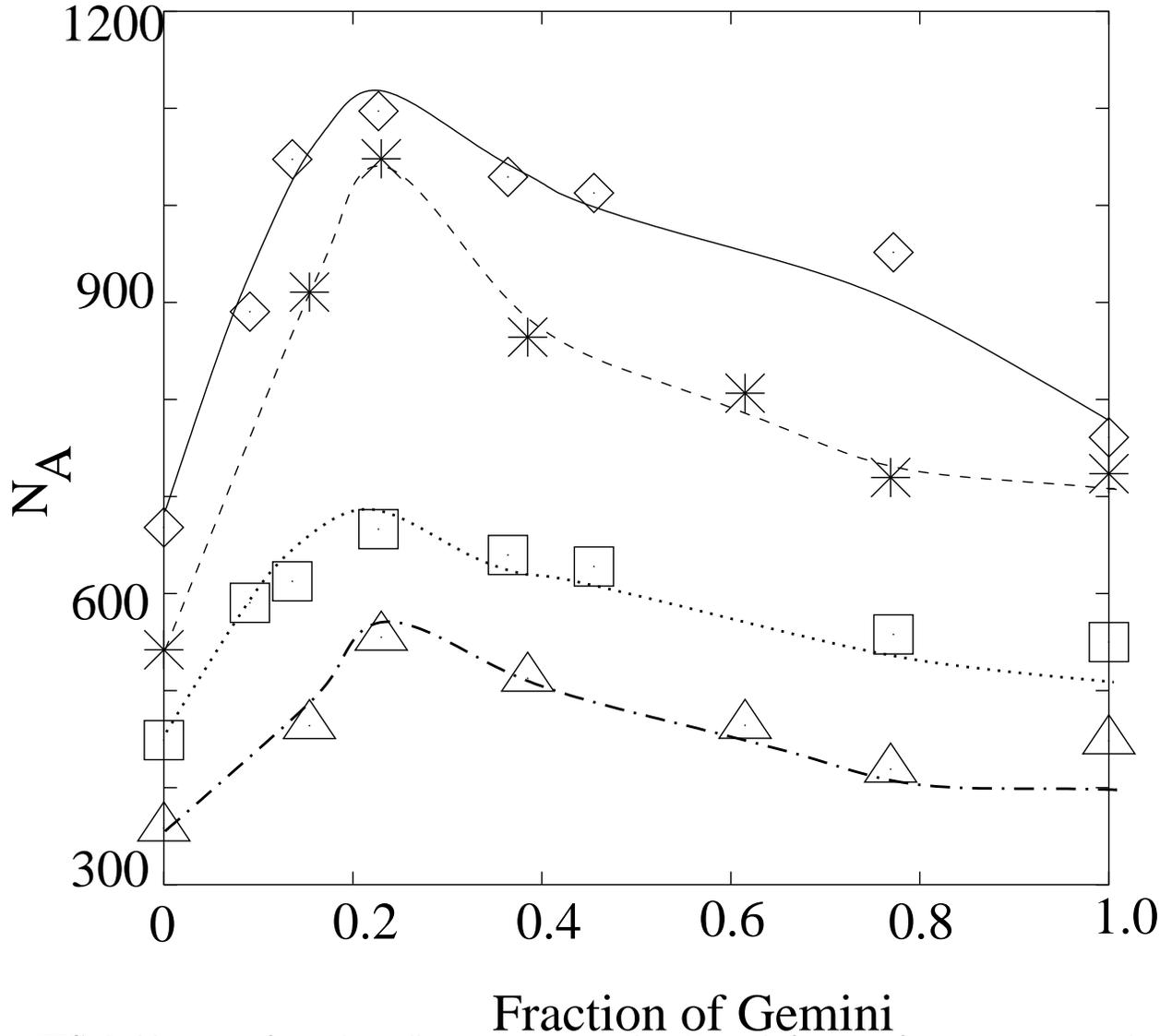}
\caption{\sf Variation of mixed micelle aggregation number $N_A$ with
fraction of gemini present in the mixture
of single-tail and gemini surfactants in water for different temperature $T$
and surfactant concentration $\phi$. The symbols
correspond to $T = 2.2 (\square) $ and $3.0 (\diamond)$ at $\phi = 0.022$ and
and $T = 2.2 (\triangle)$ and $3.0
(\ast)$ at $\phi = 0.013$.}
\label{mixagg}
\end{figure}

\begin{figure}
\epsfxsize=\columnwidth\epsfbox{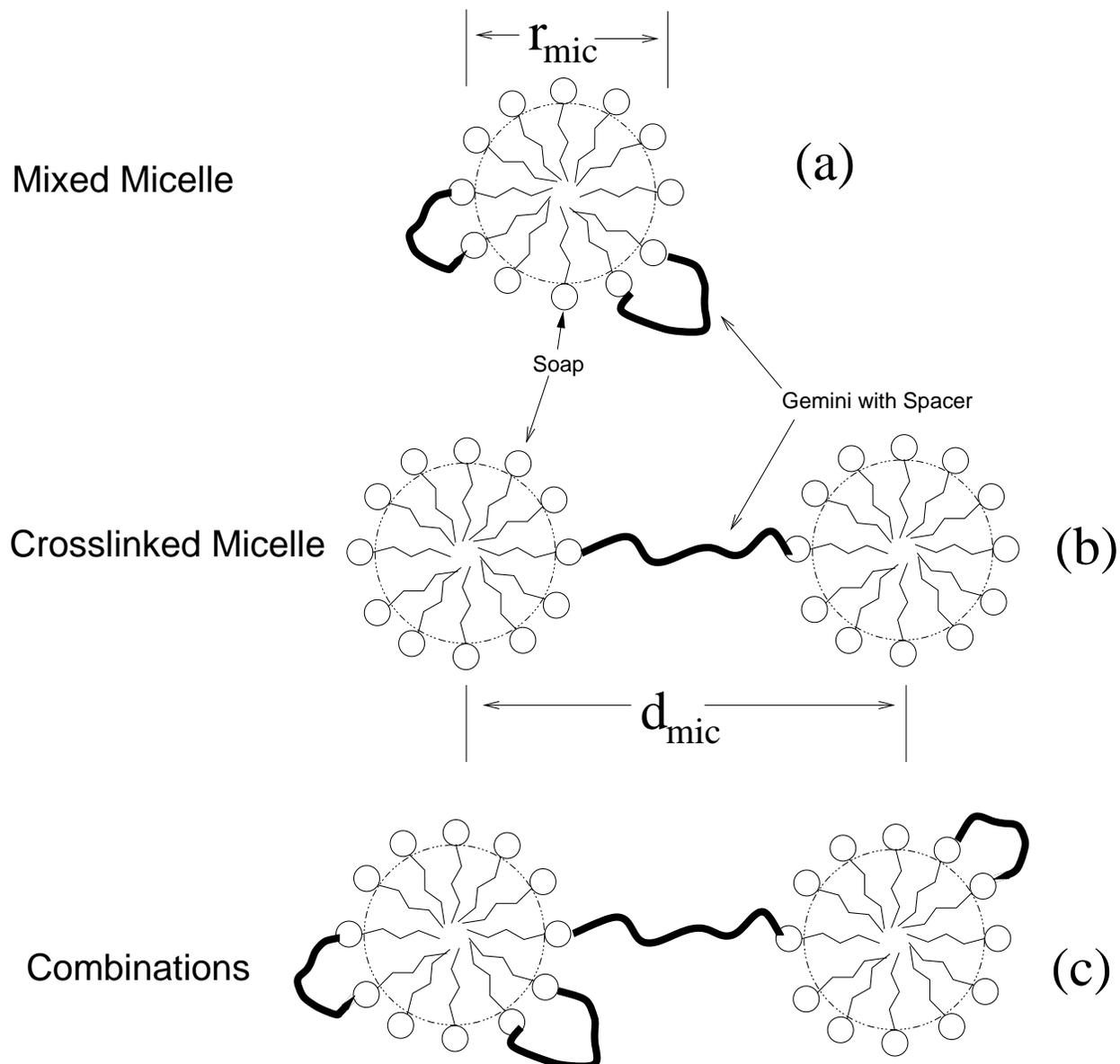}
\caption{\sf Schematic representation of a few possible supra-molecular
aggregates formed by single-tail and gemini surfactants in water.
$r_{mic}$ is the size of the micelle and $d_{mic}$ is the distance between
the micelles.}
\label{schematic}
\end{figure}

\begin{figure}
\epsfxsize=\columnwidth\epsfbox{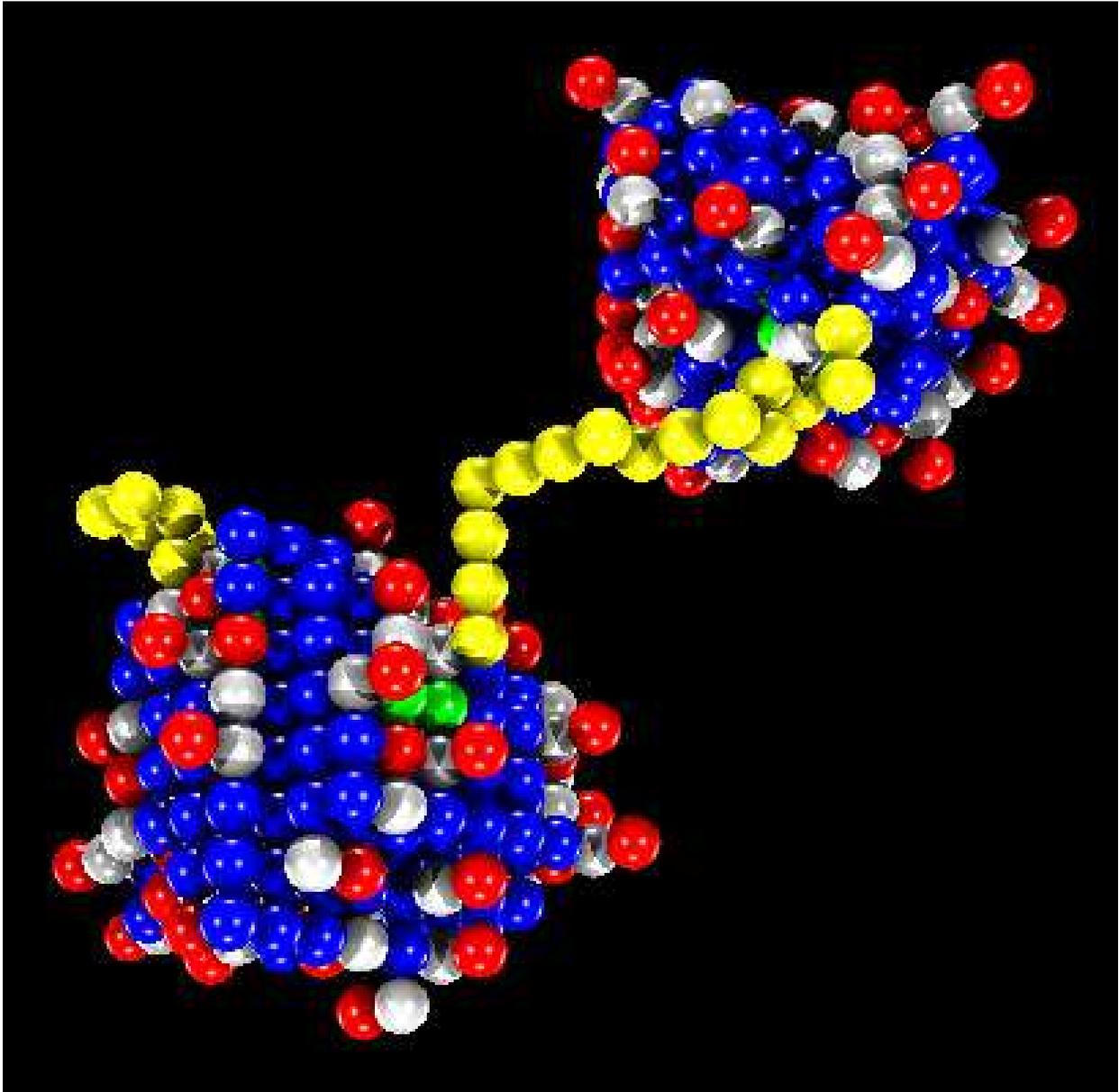}
\caption{One of the cross-linked micelles in the system is shown. 
The parameters
are: $m = 4$ for both single-tail surfactants and geminis and $n = 13$ for
the geminis while $T=2.2, \phi = 0.008$, $c_{gem}=0.1$. The monomers belonging
to the heads, neutral parts and tails of the single-tail surfactants are
represented by red, gray and blue spheres while those belonging to the heads
(as well as spacers), neutral parts and tails of the geminis are represented
by yellow, white and green spheres, respectively.}
\label{dumbbell}
\end{figure}

\begin{figure}
\epsfxsize=\columnwidth\epsfbox{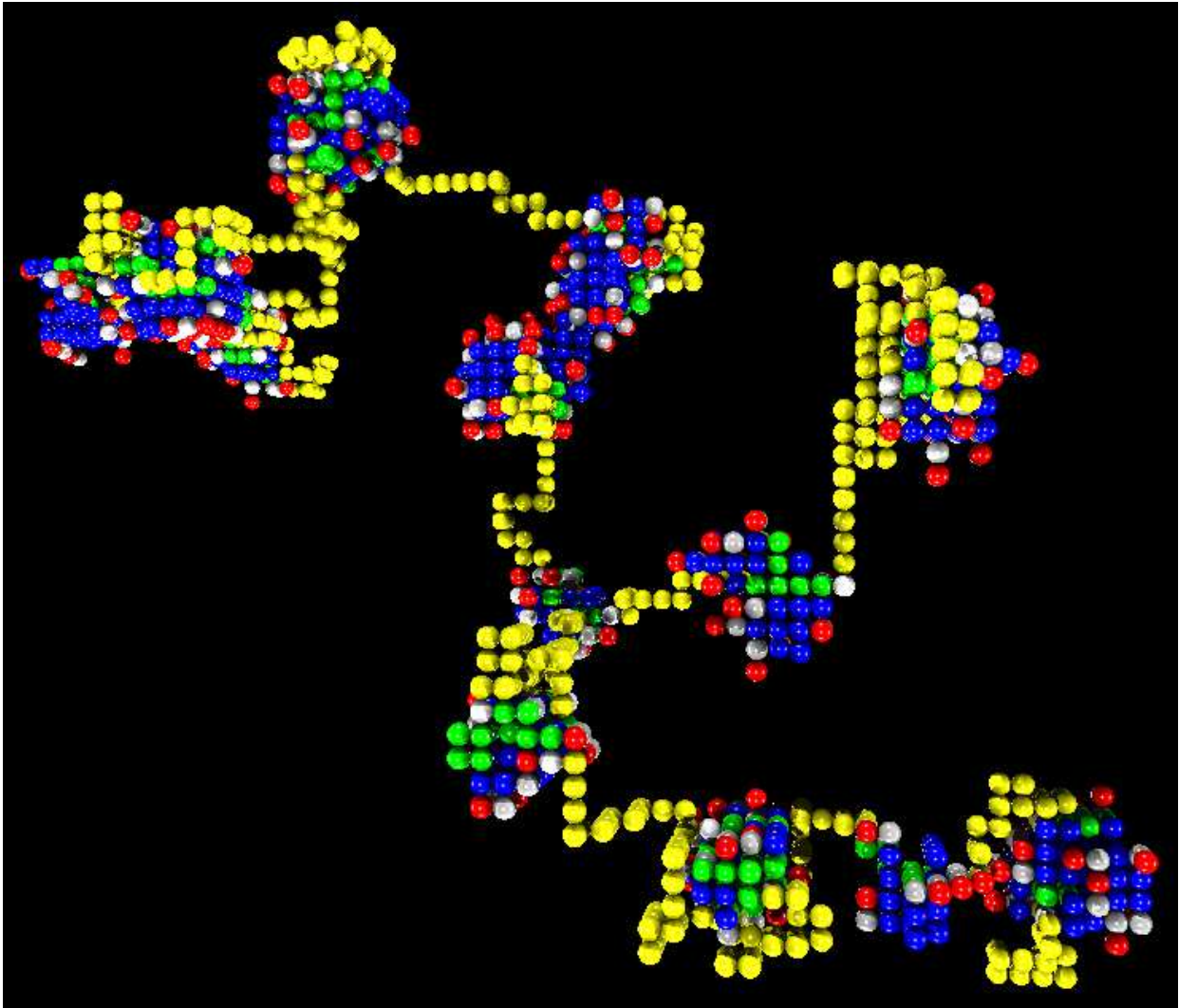}
\caption{One necklace-like aggregate (with branch)
 formed by the cross-linking of micelles in the system is
shown. The parameters are: $m=8$ for the single-tail surfactants while
$m = 4,n = 13$ for the geminis and $T=2.8, \phi = 0.008,$ $c_{gem}=0.
1$. The colour code is same as in figure ~\ref{dumbbell}.}
\label{necklace}
\end{figure}

\begin{figure}
\epsfxsize=\columnwidth\epsfbox{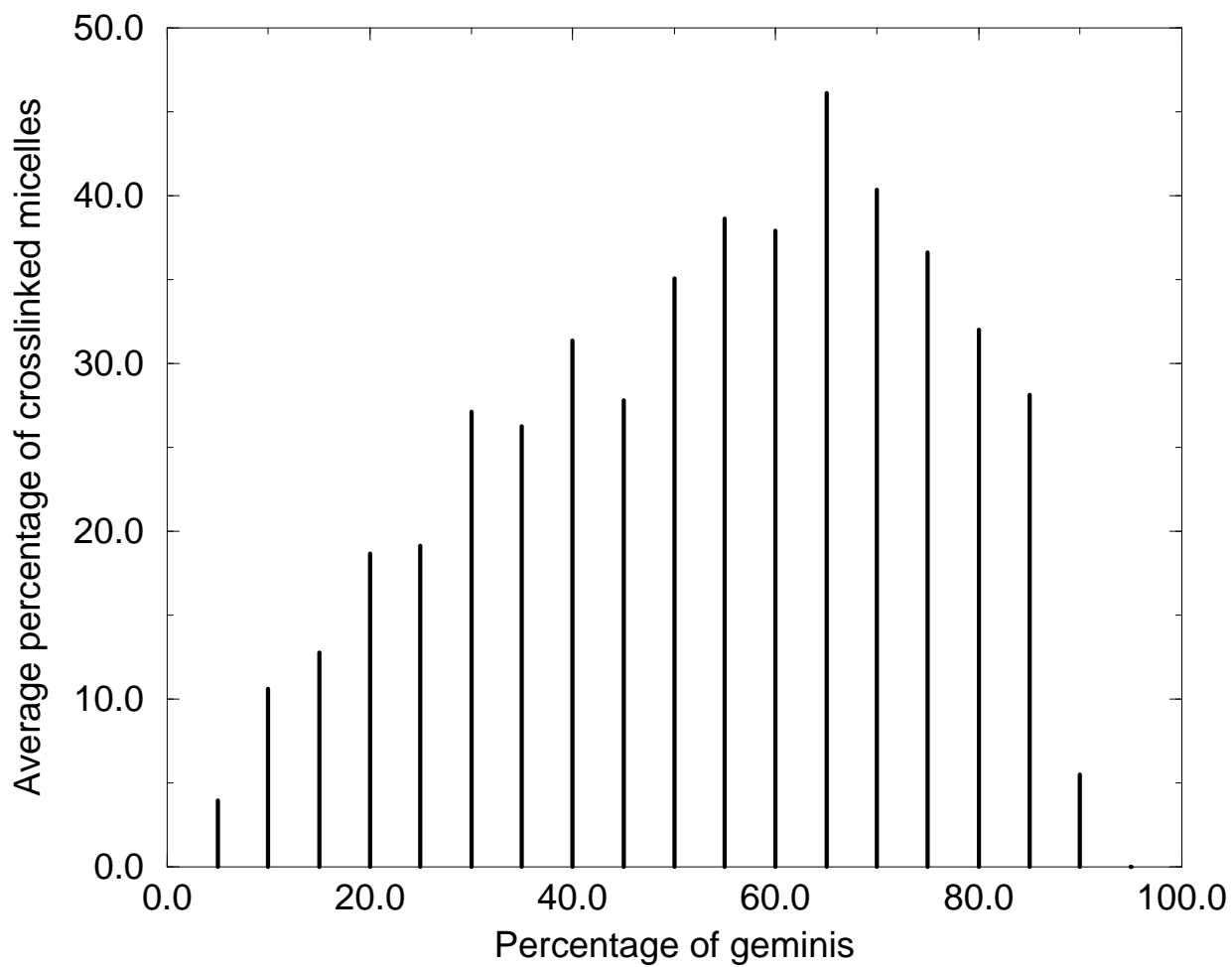}
\caption{Percentage of cross-linked micelles as a function of the
fraction of the gemini. The total concentration of the surfactants
is $0.03$ and temperature is $2.2$.}
\label{cross}
\end{figure}

\begin{figure}
\epsfxsize=\columnwidth\epsfbox{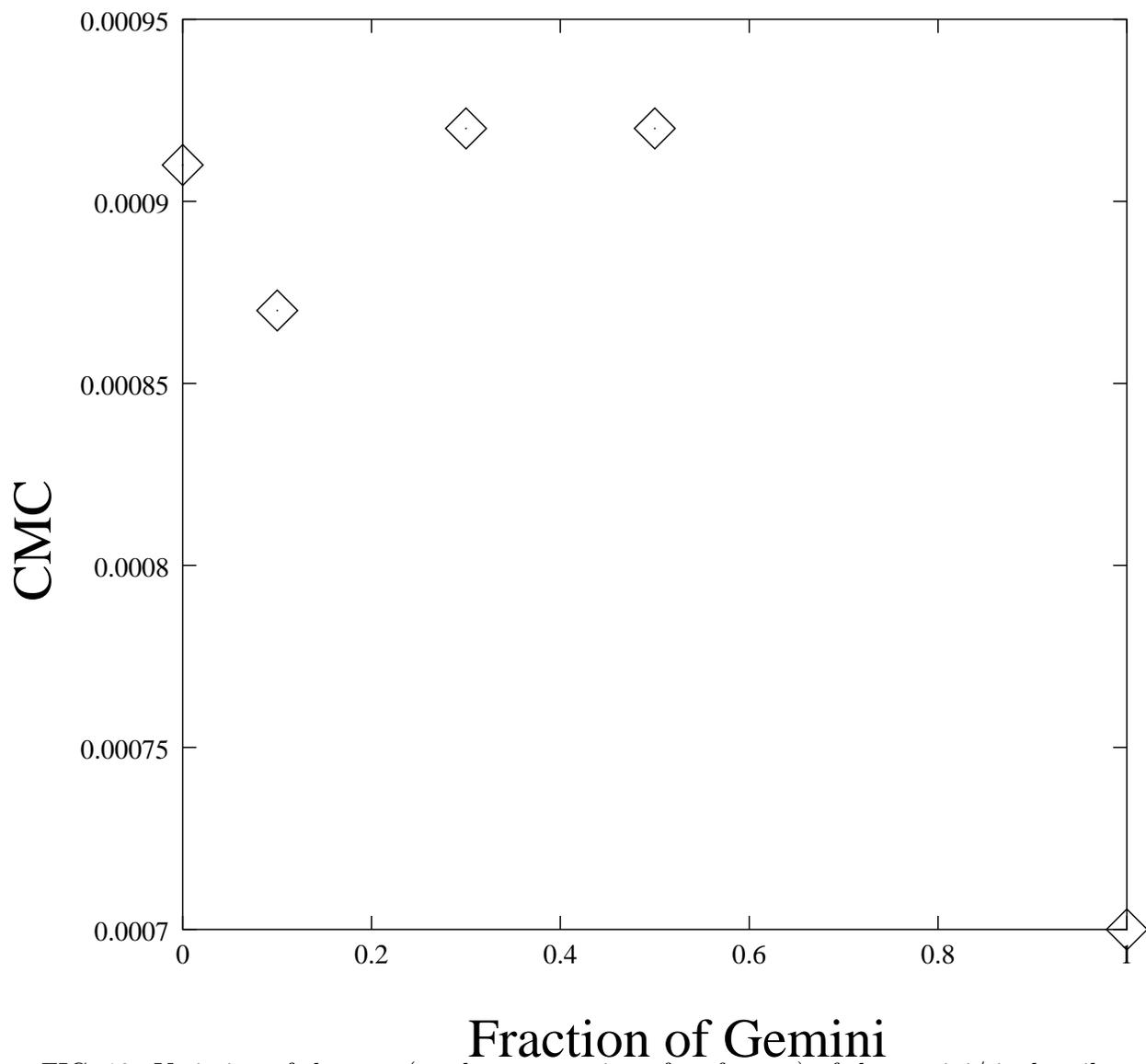}
\caption{Variation of the cmc (total concentration of surfactant) of the
gemini/single-tail surfactant mixture with gemini mole fraction at temperature
$T = 2.2 $.}
\label{cmc}
\end{figure}

\begin{figure}
\epsfxsize=\columnwidth\epsfbox{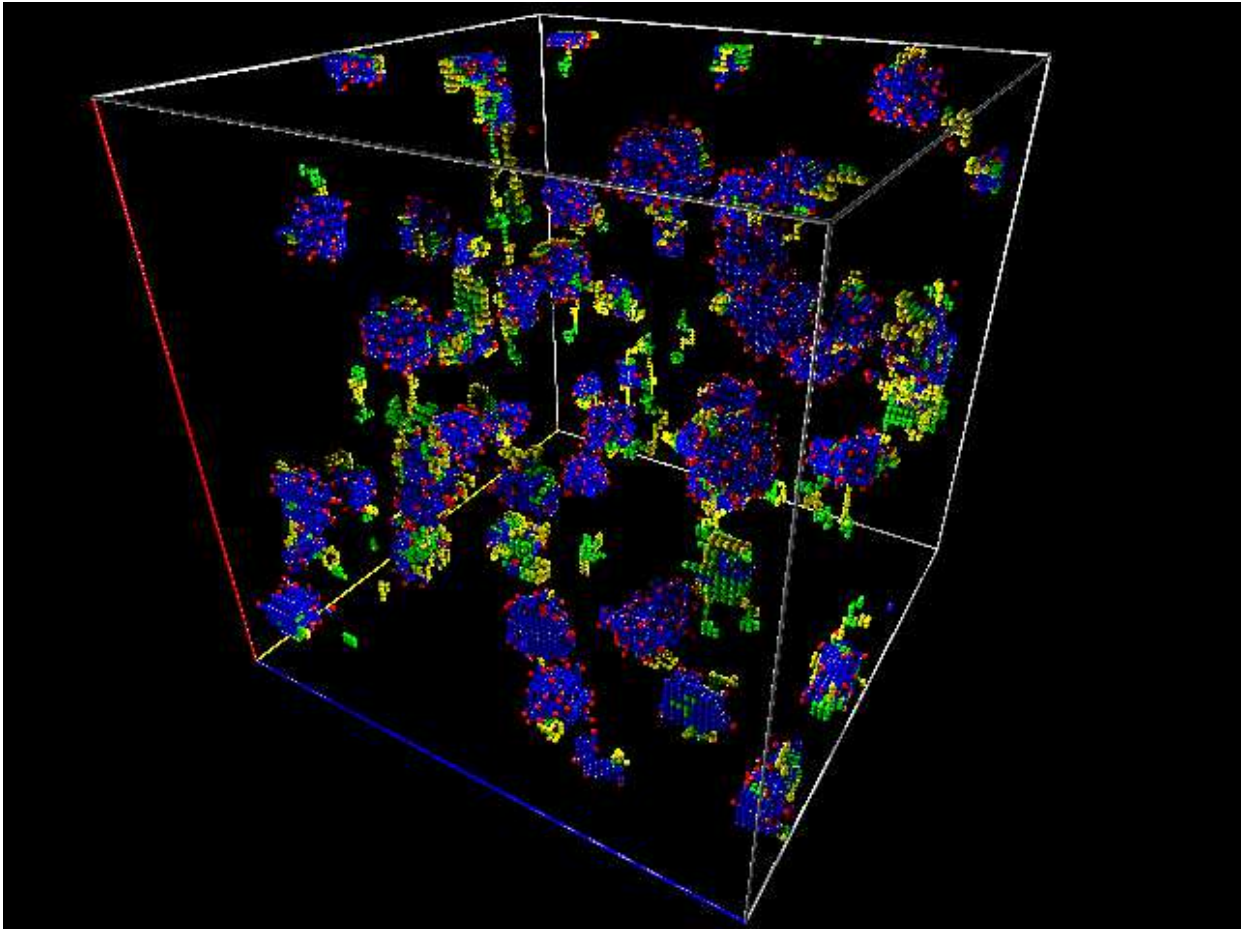}
\caption{Cross-linked micelles for hydrophobic rigid spacer
is shown. The
parameters are: $m = 10$ for both single-tail surfactants and geminis
and $n = 14$ for the geminis while $T=2.2, \phi = 0.022,$, $c_{gem}=0.005$.
The colour code is same as in figure ~\ref{dumbbell}.}
\label{bendcross}
\end{figure}

\begin{figure}
\epsfxsize=\columnwidth\epsfbox{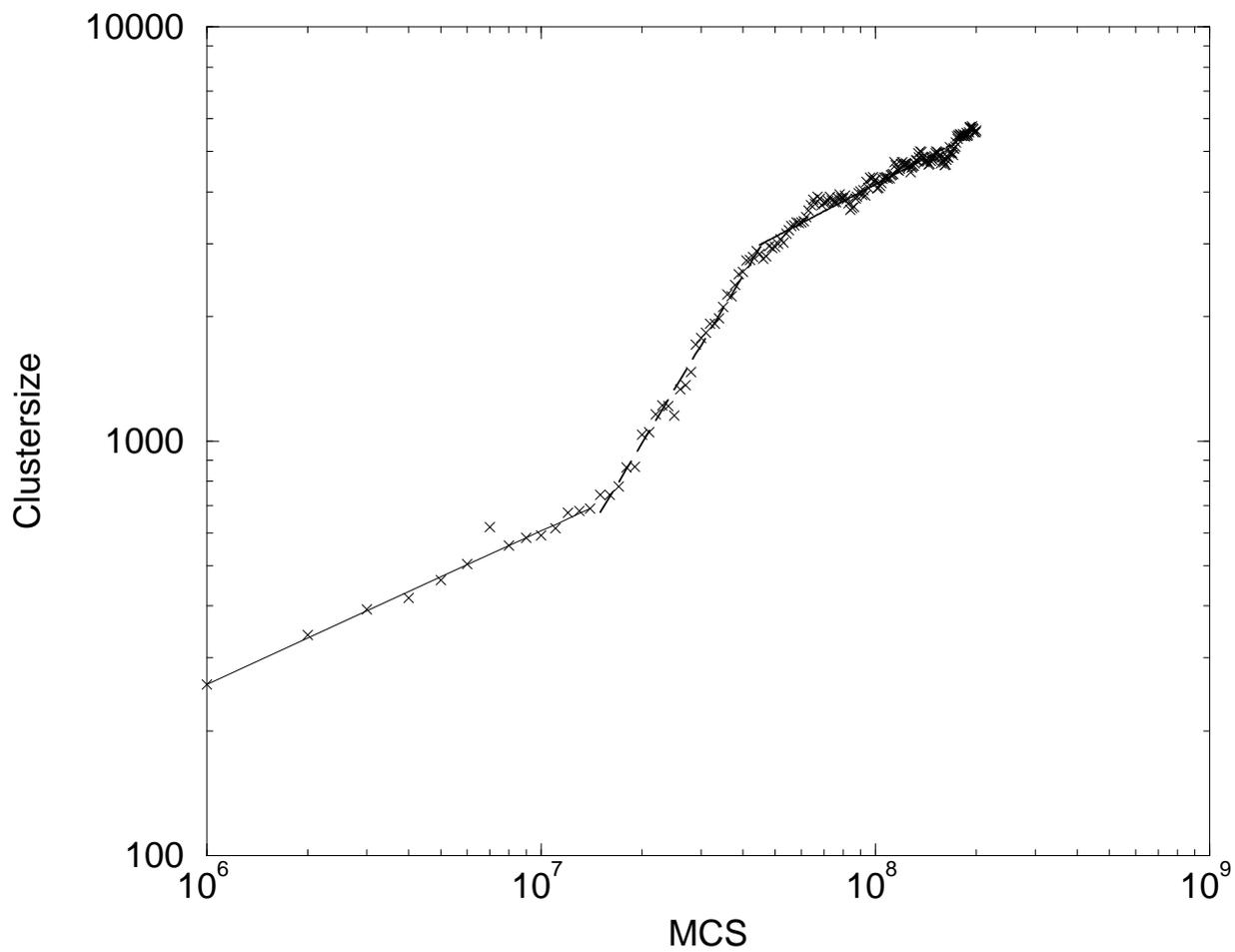}
\caption{\sf The mean-cluster size of the monomers of the 
amphiphiles is plotted against time.}
\label{growth}
\end{figure}

\begin{figure}
\epsfxsize=\columnwidth\epsfbox{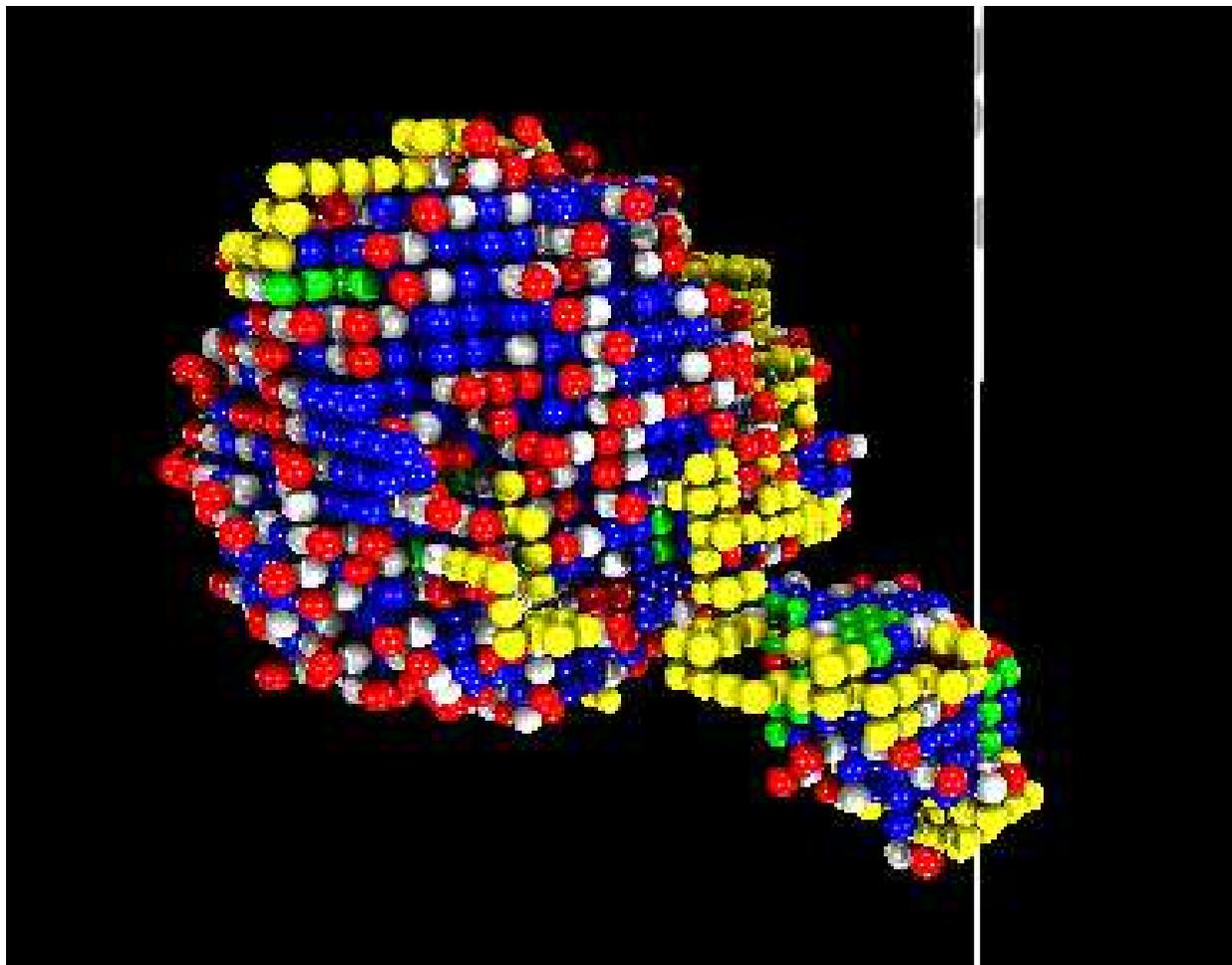}
\caption{\sf Vesicles cross-linked with micelles after $119$ million MCS.
The colour code is same as in figure \ref{dumbbell}.}
\label{vestrans}
\end{figure}

\begin{figure}
\epsfxsize=\columnwidth\epsfbox{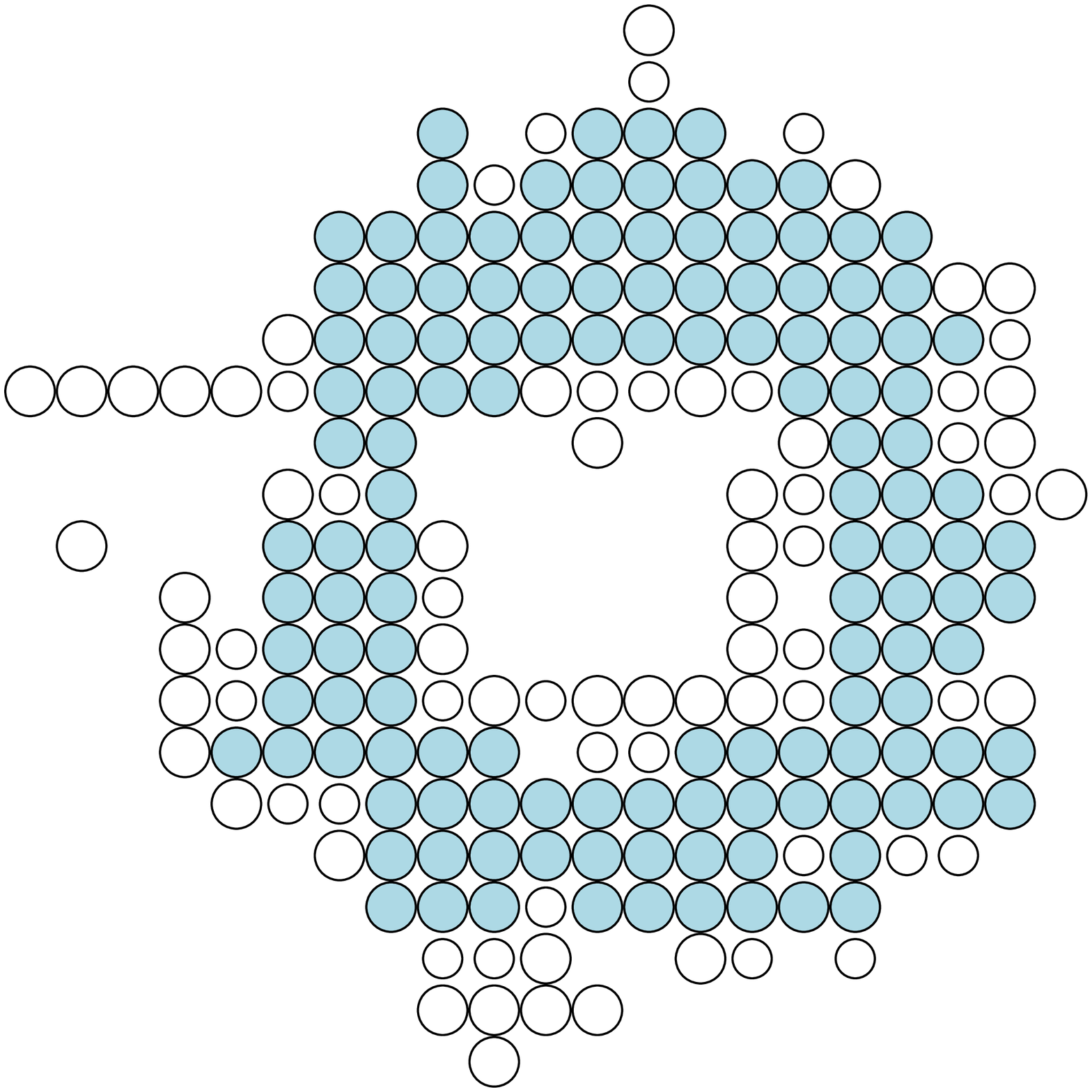}
\caption{\sf Cross-section of the vesicle.
White sphere represents hydrophilic monomers and grey
sphere represent hydrophobic monomers.}
\label{vesslice}
\end{figure}

\begin{references}
\item[$^{1}$] Present address: Dept. of Physics, 
Condensed Matter Laboratory, University of Colorado, Boulder,
CO 80309-0390, USA.
\item[$*$] Email : maiti@colorado.edu.
\item[$^{\dag}$] On leave from Physics Dept., I.I.T., Kanpur, India.
\bibitem{Evans}  Evans D. F. and Wennerstrom H., {\sl The colloidal
domain where physics, chemistry, biology and technology meet},
VCH, New York (1994).
\bibitem{Deinega} Deinega Y. F., Ulberg Z. R., Marochko L. G., Rudi V. P. and
Deisenko V. P., Kolloidn Zh. {\bf 36}, 649 (1974).
\bibitem{Menger1} Menger F. M. and Littau C. A., J. Am. Chem. Soc.,
{\bf113}, 1451 (1991).
\bibitem{Menger2} Menger F. M. and Littau C. A., J. Am. Chem. Soc.,
{\bf 115}, 10083 (1993).
\bibitem{Rosen1} Rosen M., Chemtech {\bf 23}, 30 (1993).
\bibitem{Zana1} Zana R., Benrraou M. and Rueff R., Langmuir, {\bf7}
,1072 (1991).
\bibitem{Zana2} Zana R. and Talmon Y., Nature, {\bf 362}, 228 (1993).
\bibitem{Zanarev} Zana R., in: {\it Novel Surfactants: Preparation, 
Applications and Biodegradability}, ed. Holmberg C (Dekker, New York, 1998) 

\bibitem{Rosen2} Song L. D. and Rosen M. J., Langmuir, {\bf 12}, 1149 (1996).

\bibitem{Gelbert} Gelbart W., Ben Shaul A. and Roux D. (eds.) {\sl Micelles,
Membranes, Microemulsions and Monolayers} (Springer, Berlin, 1994).
Tanford C.,
{\sl The Hydrophobic Effect: Formation of Micelles and Biological
Membranes}, (Wiley, New York 1980).

\bibitem{meli}Menger F. M. and Eliseev A. V., Langmuir, {\bf11}, 1855 (1995).

\bibitem{zanal1}Zana R., Levy H., Danino D., Talmon Y. and Kwetkat K.,
Langmuir, {\bf13}, 402 (1997).

\bibitem{zanal2} Zana R., Levy H. and Kwetkat K., J. Colloid Interface Sci.,
{\bf197}, 370 (1998).

\bibitem{Khalatur} Khalatur, P. G. and Khokhlov, A. R., Macromol. Theory
Simul., {\bf 5}, 877, (1996).
\bibitem{Semenov} Semenov, A. N., Joanny, J. F. and Khokhlov, A. R.,
Macromolecules, {\bf28}, 1066 (1995).

\bibitem{Stauffer} Stauffer D., Jan N., He Y., Pandey R. B., Marangoni D. G. and
       Smith-Palmer T., J. Chem. Phys. {\bf 100}, 6934 (1994);
\bibitem{stau1}
Jan N. and Stauffer D., J. de Physique I {4}, 345 (1994); D. Stauffer and D. Woermann, {\em J. Phys. II (France)}, {\bf 5}, 1 (1995).

\bibitem{Livrev}
  Liverpool T. B., in: {\sl Annual Reviews
  of Computational Physics}, vol. IV, ed. D. Stauffer (World
  Scientific, Singapore 1996). 
   Schmid, F. in : {\sl Computational methods in colloid and interface 
science}, ed. M. Borowko, (Marcel Dekker inc.), in press. 

\bibitem{bernardes} Bernardes A. T., J. Phys. II France {\bf 6}, 169 (1996);
see also Langmuir, {\bf 12}, 5763 (1996).

\bibitem{maiti} Maiti P. K. and Chowdhury D., Europhys. Lett. {\bf 41}, 183, (1998); J. Chem. Phys. {\bf 109}, 5126 (1998)

\bibitem{smit} Smit B., Hilbers P. A. J., Esselink K., Rupert L. A. M., 
van Os N. M. and Schlijper A. G., Nature, {\bf 348}, 624 (1990); 
Karaboni S., Esselink K., 
Hilbers P. A. J., Smit B., Karthauser J., van Os N. M. and Zana R., 
Science, {\bf266}, 254 (1994).

\bibitem{Lar1}
Larson R. G., Scriven L. E. and Davis H. T., J. Chem. Phys. {\bf83}, 2411
 (1985); Larson R. G., J. Chem. Phys. {\bf89}, 1642 (1988) and
{\bf 91}, 2479 (1989).

\bibitem{Gomsch} Gompper G. and Schick M.,
{\sl Phase Transitions and Critical Phenomena}, Vol. 16,
ed. C. Domb and J. L. Lebowitz, (Academic Press, London 1994).

\bibitem{Chow2} Chowdhury D., Langmuir, {\bf12}, 1098 (1996);
Chowdhury D., Maiti P. K., Sabhapandit S. and Taneja P., Phys. Rev.E, {\bf 56},
 667 (1997).

\bibitem{binder} Binder K. (ed.) {\it Applications of the Monte Carlo Method in 
Statistical Physics}, Topics in Applied Physics 36, (Springer, 1987).

\bibitem{hoshen} Hoshen J. and Kopelman R., Phys. Rev. B{\bf14}, 3438 (1976);
see also Stauffer D. and Aharonmy A., {\it Introduction to percolation 
therory}, (Taylor \& Francis, London, 1992).

\bibitem{flimm} Flimm O., Masters Thesis, University of Cologne
(unpublished, 1999)

\bibitem{zanaold} Zana R. and Weill C., J. Phys. Lett. {\bf46},L953 (1985).

\bibitem{Rgao}Rosen M. J.,  Gao T., Nakatsuji Y. and Masuyamai A., Colloids
Surf. A: Physicochem. Eng. Asp. {\bf88}, 1 (1994).

\bibitem{Rzhu} Rosen M. J., Zhu Z. and  Gao T., J. Colloid Interface Sci.,
{\bf15}, 224 (1993).

\bibitem{gunton} Gunton J. D. and Droz M., {\sl Lec. Notes in Physics}, 
vol. 183 (1983) ; A.J. Bray, Adv. Phys.{\bf 43}, 357 (1994).

\end{references}
\end{document}